\documentclass[a4paper]{article} 
\usepackage{epsfig}
\usepackage{amsmath}
\usepackage[left=2cm,top=2cm,bottom=3cm]{geometry}

\makeatletter 
\@addtoreset{equation}{section} 
\makeatother 
 
\newcommand{\be}{\begin{equation}} 
\newcommand{\ee}{\end{equation}} 
\newcommand{\bea}{\begin{eqnarray}}
 \newcommand{\eea}{\end{eqnarray}}
 \def\non{\nonumber }
\newcommand{\ov } {\over } 
\newcommand{\p }{\partial }
 \newcommand{\s }{\sigma }
 
\def\r{r }

\def\td{\tilde }
\def\a{\alpha }

\def\w{\omega }
 
\def\appendix#1{   \addtocounter{section}{1}   \setcounter{equation}{0}   
\renewcommand{\thesection}{\Alph{section}}   \section*{Appendix \thesection\protect\indent \parbox[t]{11.15cm}   {#1} }   \addcontentsline{toc}{section}{Appendix \thesection\ \ \ #1}   } 

\def\appendix#1{
  \addtocounter{section}{1}
  \setcounter{equation}{0}
  \renewcommand{\thesection}{\Alph{section}}
  \section*{Appendix \thesection\protect\indent \parbox[t]{11.15cm}
  {#1} }
  \addcontentsline{toc}{section}{Appendix \thesection\ \ \ #1}
  }

\def \td {\tilde} 

\begin{document}

 %\vspace*{1cm}
\null\vskip-24pt 
   \hfill UB-ECM-PF/07/14
      \vskip-1pt
\hfill {\tt hep-th/yymmnnn}
\vskip 1truecm
\begin{center}
\vskip 0.2truecm {\Large\bf Strings and D-branes in a supersymmetric magnetic \\
flux background
 }
\vskip 0.2truecm

\vskip 0.7truecm
\vskip 0.7truecm

{\bf Roberto Iengo$^a$, Jaume Lopez Carballo$^b$ and Jorge G. Russo$^{b,c}$}\\
\vskip 0.4truecm
\vskip 0.4truecm

${}^a${\it  International School for Advanced Studies (SISSA)\\
Via Beirut 2-4, I-34013 Trieste, Italy} \\
{\it  INFN, Sezione di Trieste}

\medskip

$^{b}${\it 
Departament ECM,
Facultat de F\'\i sica, Universitat de Barcelona,
 Spain} 
 
 \medskip

$^{c}${\it 
Instituci\' o Catalana de Recerca i Estudis Avan\c{c}ats (ICREA)
} 
 
\end{center}
\vskip 0.2truecm 

\noindent\centerline{\bf Abstract}

We investigate how the presence of RR magnetic $F_{p+2}$ fluxes affects
the energy of classical Dp branes,
 for specific string theory supersymmetric backgrounds which are solutions to the leading order in $\a' $  including
back-reaction effects. The Dp brane dynamics is found to be similar to the 
well known dynamics of particles and strings moving in magnetic fields. 
We find a class of BPS solutions
which generalize the BPS fundamental strings or BPS branes with momentum and winding to the
case of non-zero magnetic fields. Remarkably, the interaction with the magnetic fields does not
spoil the supersymmetry of the solution, which turns out to be invariant under four supersymmetry transformations.
We find that magnetic fields can significantly reduce  the energy of some BPS 
strings and Dp branes, in particular, some macroscopic Dp branes become 
light for sufficiently large magnetic fields.

\newpage

\renewcommand{\thefootnote}{\arabic{footnote}}
\setcounter{footnote}{0}

%%%%%%%%%%%%%%%%%%%%%%%%%%%
\section{  Introduction}
%%%%%%%%%%%%%%%%%%%%%%%%%%%%%%%%%%%%%%%%%%%%%%%%%%%%%%%%%%%%%%%%%%%%%%%%%

\setcounter{equation}{0}

Flux compactifications in string theory stabilize the moduli and presently provide the main framework to
connect strings to the real world \cite{Douglas,Denef}. The fluxes induce a non-trivial  
potential energy for the moduli. In addition, the fluxes can significantly affect  the energy spectrum of 
some quantum states. 
The precise modification of the energy of a given quantum state 
 depends on the particular string compactification, but there 
are some general features which are part of the well known physics of particles interacting with magnetic fields.
In quantum theory, the energy of a state with  electric charge $e$ and spin $S$ moving
in a magnetic field is given by the formula
\be
E^2= k_i^2 + 2 e B_z (l+{1\over 2} - S_z) \ ,\qquad l=0,1,2,...\ ,\ \ \ -S\leq S_Z\leq S\ .
\ee
For states with spin aligned with the magnetic fields, the energy is reduced and the state can even become tachyonic
for certain values of the magnetic field, leading to instabilities.

A natural question is whether analogous instabilities could be  present in superstring theory in spaces
containing NSNS or RR magnetic fluxes turned on in some directions.
In the context of type II string theory, there are exactly solvable models where the magnetic
field originates by Kaluza-Klein reduction from the metric or from the antisymmetric tensor $B_{\mu\nu}$
\cite{RTmagnetic}. 
These models include gravitational back reaction effects due to the energy density of the magnetic field, and
exhibit tachyonic instabilities for magnetic fields greater than some  critical values
\cite{RTmagnetic,RTinstab,David}.

It is possible to consider similar closed string models with magnetic field configurations that preserve supersymmetry
\cite{RTSS}.
In this case the physical string spectrum is tachyon free. Nevertheless, one can expect that for certain values of the magnetic field, the energy can be significantly reduced due to a negative contribution from the 
gyromagnetic interaction, so that a macroscopic string of size $\gg l_s$
can even become light, $E\sim l_s^{-1}$. 
In this paper we will show that this is indeed the case and study the analog phenomenon for Dp branes. 
To this aim, we will consider a string-theory background with a RR magnetic $F_{p+2}$ flux which is obtained by S and T-dualities from the  magnetic background found in \cite{RTSS}.

This background has two magnetic parameters $B_1$, $B_2$. When $B_1=B_2$, the background preserves 1/2 of the 32 supersymmetries. We will find classical string and 
Dp brane configurations becoming light for certain values of the magnetic 
field parameters $B_1, B_2$. Remarkably, these solutions are BPS, despite the interaction with the magnetic field.
The solutions break an additional fraction 1/4 of the 16 supersymmetries of the background and therefore are invariant under  4 supersymmetry transformations. 

Studies of  Dp brane classical solutions in flat space can be found for example
in \cite{Hoppe} and more recently \cite{Axenides}. Different studies of the conditions to have supersymmetric Dp branes in some supersymmetric compactifications with Ramond-Ramond (RR) fluxes are in \cite{Frey,Gomis,Martucci}.

The organization of this paper is as follows. In Sect. 2.1 we review the string spectrum 
in a particular flat (but globally non-trivial) background which gives rise to a  magnetic field 
by Kaluza-Klein reduction; we recall the supersymmetry properties of the background and
the presence of tachyons when supersymmetry is  broken (the main points of this quantum analysis are 
reviewed in the Appendix A).  
In Sect. 2.1.1  we identify  BPS states 
for the supersymmetric background with $B_1=B_2$.
 In Sect. 2.2 we construct a family of classical solutions which corresponds to BPS states by solving the classical equations of motion. 
In Sect. 2.3 we show that these classical solutions indeed preserve a fraction of 
supersymmetry, by using both cartesian and polar coordinates.
In Sect. 3.1 we study the  background obtained by a T-duality transformation; this background contains 
explicitly a $B_{\mu\nu}$ field and the metric is curved due to the back reaction produced by the magnetic energy density. The spectrum, and in particular the BPS states, are obtained from  Sect. 2.1 by the standard T-duality rules. In Sect. 3.2 we consider a family of classical solutions corresponding to the BPS states 
by solving the classical equations in the background of Sect. 3.1 by using the Polyakov formalism.
In Sect. 3.3  we repeat this computation using the Nambu-Goto formalism; this is done in preparation for the study of classical Dp-branes solutions, whose dynamics is  governed 
by the Dirac-Born-Infeld  action.
In fact, in the case of the Dp-branes, studied in Sect.4, the method and the
various steps for obtaining the solutions will be seen to be essentially the same. 
%Here the background is obtained via S-duality and various T-dualities from the background of Sect. 3.1. 
We find a family of BPS classical solutions, 
the energy spectrum being a  generalization of the BPS spectrum in the magnetic background discussed in Sect. 3. 
One important feature of the result is the presence of light states which have macroscopic
features. This result is further discussed in Sect. 5. 
Finally, in the Appendix B we consider  another family of classical F-string 
solutions in the background of Sect. 3 (or, equivalently, D-string  solutions in the background of 
Sect. 4). These solutions, although  not BPS, are nevertheless interesting because they generalize 
to a non-flat background with fluxes the widely studied case of the folded rotating string.

%%%%%%%%%%%%%%%%%%%%%%%%%%%%%%%%%%%%%%%%%%%%%%%%%%%%%%%%%%%%%%%%%%%%%%%%%
\section{  Fundamental string in magnetic backgrounds}
%%%%%%%%%%%%%%%%%%%%%%%%%%%%%%%%%%%%%%%%%%%%%%%%%%%%%%%%%%%%%%%%%%%%%%%%%
\setcounter{equation}{0}
%%%%%%%%%%%%%%%%%%%%%%%%%%%%%%%%%%%%%%%%%%%%%%%%%%%%%%%%%%%%%%%%%%%%%%%%%
\subsection{Magnetic fields from Kaluza-Klein reduction}
%%%%%%%%%%%%%%%%%%%%%%%%%%%%%%%%%%%%%%%%%%%%%%%%%%%%%%%%%%%%%%%%%%%%%%%%%

We shall consider a simple magnetic string model  given in terms of the background \cite{RTSS}
\be
ds^2 = -dt^2+dx_s^2+dy^2+ dr_1^2+r_1^2 (d\varphi_1 + B_1 dy)^2+dr_2^2+r_2^2 (d\varphi_2 + B_2 dy)^2 \ ,
\label{polar}
\ee
where all other supergravity fields are trivial and $s=6,...,9$.
This is an exact conformal string model, since the metric is flat. 
It is globally non-trivial, due to the fact that $y$ is a periodic coordinate, $y=y+ 2\pi R$.

The background preserves 1/2 of the 32 supersymmetries provided 
\be
B_1=\pm B_2 \ .
\ee
The string model is a simple generalization of the non-supersymmetric string model 
with $B_2=0$ that was  solved in \cite{RTmagnetic}. 
The exact physical string spectrum is found in a similar way and it is given by \cite{RTSS}
\begin{equation}
\begin{aligned}
\a ' M^2 & = 2(N_L+N_R) +\a' ({n\over R}- B_1 J_1-B_2J_2)^2+ {m^2R^2\over\a' }
\\
&\quad - 2 B_1Rm (J_{1R}-J_{1L}) - 2 B_2Rm (J_{2R}-J_{2L})  \ ,
\\
N_R- N_L & = m n \ ,
\label{qqqq}
\end{aligned}
\end{equation}
where $B_1$ and $B_2$ are in the interval $0\leq \gamma_{1,2}  <1 $, $\gamma_{1,2} \equiv B_{1,2}Rm$. For other intervals the spectrum is repeated periodically in the parameters $\gamma_{1,2} $ with period 1. The angular momentum operators for the two planes
$(r_1,\varphi_1)$ and $(r_2,\varphi_2)$ are given by 
\be
J=J_R+J_L\ , \qquad J_{L,R}=\pm \bigg( l_{L,R} +{1\over 2}\bigg)+S_{L,R}\ ,
\ee
where for the sake of clarity we omitted the obvious indices 1,2.
$N_{L,R}=0,1,2,.... $ are the standard excitation number operators of flat-space type II superstring theory
and $S_{L,R}$  are the standard Left and Right contributions to the flat-space spin operator (the main points
of the derivation are reported in the Appendix A, see \cite{RTmagnetic} for details).
The parameters $l_{L,R}=0,1,2,...$  are orbital angular momenta (Landau numbers).
The parameters $m,n$ represent winding and momentum in the $y$ direction. If there are other compact coordinates among the $x_s$, then their winding and momentum contributions to the energy is added in the standard way as 
in the flat case. The spin operators satisfy the inequalities
\be
\big|S_{1L,R}+ S_{2L,R}\big|\leq N_{L,R}+1 \ ,\qquad \big|S_{1L,R}- S_{2L,R}\big|\leq N_{L,R}+1 \ .
\ee
One finds that $M^2\geq 0$ for $B_1=\pm B_2$ (see Appendix A), but
The spectrum contains tachyons  in some regions of the parameter space \cite{RTmagnetic,RTSS},
if $B_1\neq \pm B_2$.
For example, consider a state with the following quantum numbers:
\be
\begin{aligned}
N_R & =N_L=0\ ,& S_{1R}&=-S_{1L}=1\ ,& S_{2R} & =S_{2L}=0\ ,  \\ l_{1L,R} & =l_{2L,R}=0\ ,&  m & =1 \ , &  n& = 0\ ,  
\end{aligned}
\label{tqq}
\ee
so that $J_1=J_2=0\ ,\ J_{1R}-J_{1L}=1\ ,\ J_{2R}-J_{2L}=-1$ and 
\be
\a' M^2 = {R^2\over \a' } -2 (B_1-B_2)R  \ .
\ee
The state becomes tachyonic for $B_1-B_2> R/(2\a ')$.
In the case $B_1=B_2$, the full spectrum is tachyon free for any $B_1$, consistently with supersymmetry.

Another example is a state classically identical to the BPS state discussed below,
see Sects.2.1.1 and 2.2. In Appendix A we show that, for instance for $B_2=0$, this state can have $M^2<0$
in some parameters range. For $B_2=B_1$ (taking here both positive) the positive zero point energy of 
the Landau levels in the plane $2$ (a quantum effect of the sigma model) insures that $M^2\geq 0$.

Therefore the classical analysis, that would give $M^2$ as a sum of contributions $\geq 0$,
is not enough to guaranty the absence of tachyons, if supersymmetry is completely broken. 
However, this occurs when in parameters regions where the classical result for the mass is
microscopic, that is of the order of the inverse string length or of the inverse compactification radius.

Most of the present paper is devoted to the study of states having macroscopic features but
microscopic mass. Those are precisely the cases which could give rise to tachyonic instabilities, 
if they are not protected by supersymmetry.  
 
\smallskip

In \cite{taka,Dudas} the D brane spectra in the background (\ref{polar}) has been determined by using the boundary
state formalism,  exhibiting a number of interesting features (the orientifold spectrum was studied in \cite{Angelan}). In section 4, instead,  we will be  interested in studying Dp branes not in (\ref{polar}), but in backgrounds containing $F_{p+2}$ fluxes, which are obtained from (\ref{polar}) by dualities. The spectrum, and the physics  in general, are clearly different in each case, the former \cite{taka,Dudas}  being analogous to a neutral particle (or zero-winding string) moving in the geometry (\ref{polar}) (and on the curved-space generalization discussed in section 3),
whereas the latter (section 4) involves a physics  which is similar to that of a charged particle moving in a magnetic field.

%%%%%%%%%%%%%%%%%%%%%%%%%%%%%%%%%%
\subsubsection{BPS states}
%%%%%%%%%%%%%%%%%%%%%%%%%%%%%%%%%%

In the  $B_1=B_2=0$ case, an important class of quantum string states are the BPS
string states  with \cite{DH}
\be
N_L=0\  ,\qquad N_R=mn\ .
\label{bpss}
\ee
Then $M^2 $ becomes a perfect square:
\be
\a' M^2_{\rm BPS} = 2 mn + \a'{n^2\over R^2} +{m^2R^2\over\a' } =\a' \left({n\over R} +{mR\over \a'}\right)^{\!\!2} \ ,
\ee
or
\be
M_{\rm BPS}= \left| {n\over R} + {mR\over\a'} \right| \ .
\ee
Note that $ mn =N_R>0$. When $mn <0$,  there are BPS states with $N_R=0$, $N_L=- mn$ for which 
\be
M_{\rm BPS} = \left| {n\over R} - {mR\over\a'}\right|= \left|{n\over R}\right| + \left|{mR\over\a'}\right|\ .
\ee

\medskip

We now look for similar supersymmetric states in the presence of  magnetic fields $B_1, B_2$. 
We will restrict to the case $B_1=B_2\equiv B$ where the background preserves 16 supersymmetries.
The state~(\ref{bpss}) has the energy 
\be
\begin{aligned}
\a ' M^2 &=  2mn +\a'{n^2\over R^2} +{m^2R^2\over\a' } - 2B \a' {n\over R}  (J_1+ J_2)-2 B mR (J_{1R}+J_{2R}-J_{1L}
-J_{2L}) \\
&\quad + \alpha' B^2 (J_1+J_2)^2 \ .
\end{aligned}
\ee
This becomes a perfect square if
$J_1+J_2=J_{1R}+J_{2R}-J_{1L}
-J_{2L} $, i.e.,
\be
S_{1L}+S_{2L}=-1-l_{1L}-l_{2L} \ .
\ee
Since $N_L=0$, $S_{1,2L}$ can be zero or $\pm 1$ and $|S_{1L}+S_{2L}|\leq 1$. This implies $l_{1L}=l_{2L}=0$ and
the following possible values:
\be
(S_{1L},S_{2L})=(-1,0)\ \ {\rm or} \ \ (0,-1)\ .
\label{qcond}
\ee
Then we get 
\be
M^2_{\rm BPS} = \left( {n\over R} + {mR\over\a'}- B (J_1+J_2) \right)^{\!\!2} \ .
\label{saz}
\ee
Remarkably, the
mass square is  a perfect square which indicates that the state is still supersymmetric
despite the presence of the magnetic fields. It is worth noting that 
in the case $B_1\neq B_2$, $M^2$  is not a perfect square.
Moreover, the state can become tachyonic above some critical magnetic field.

Assume for definiteness $B>0$, $m,n>0$. Taking into account the fact that $mRB<1$ and $(J_1+J_2)\leq N_R=mn $,
one can see that ${n\over R}  - B(J_1+J_2)>0$ and therefore
eq.~(\ref{saz}) can be written as 
\be
M_{\rm BPS}=  {|m|R\over\a'}+ \left|{n\over R}  - B(J_1+J_2)\right| \ .
\label{eebb}
\ee
This form will be suitable for comparison with the energy of classical strings.

\medskip

The effect of the magnetic field on a quantum state with spin aligned with the magnetic field
is to reduce its energy.
The maximum reduction is attained for the state with maximum spin $S_{1R}+S_{2R}=N_R+1$
and $l_{1,2R}=0$, so that $J_1+J_2= N_R= nm$.
{} For this state
\be
M_{\rm BPS}=  {|m|R\over\a'}+ \left|{n\over R}  - Bmn\right| \ .
\ee
The minimum mass occurs when $B$ approaches the limit of the interval, $BRm \to 1$, where
\be
M_{\rm BPS}\to  {|m| R\over \alpha '} \ .
\label{oop}
\ee
Assuming that $R$ is of the same order of magnitude as $\a' $, 
this  implies  that there are states with very large $n$ and small $m$
which become light (mass of $O(R/\a' )$) at some magnetic field. Such states have a macroscopic mass 
$ = O(n/R)\gg {1\over\sqrt{\a'}} $ in the absence of magnetic fields. 

This result can be understood from the periodicity of the spectrum: the full quantum string spectrum at $BRm\to 1$ is the same as the spectrum at $B=0$, with some relabelling of the states.
In the coordinate system where the metric becomes Minkowski at $B=0$, a given large classical string  becomes very light, $E\sim 1/l_s$, as $BRm$ approaches 1. But when $BRm\to 1$ the metric approaches the Minkowski metric 
in another coordinate system where  the polar coordinate is
$\varphi_{1,2}' =\varphi_{1,2} +y/mR $. 
The energy (\ref{oop}) then corresponds to a state with $n'=0$ and orbital angular momentum (see sect 2.2).

%%%%%%%%%%%%%%%%%%%%%%%%%%%%
\subsection{Classical BPS solution}
%%%%%%%%%%%%%%%%%%%%%%%%%%%%%%%%%%%%

There is an exponential number $=O\big( \exp(2\sqrt{2}\pi\sqrt{N_R}) \big)$
of quantum states satisfying the condition $N_R=mn $, $N_L=0$.
A particular class of classical solutions representing a small subset of these states is given by 
\be
\begin{aligned}
Z_1&=r_1 e^{i\varphi_1}= X_1+iX_2 = L_1\ e^{i k_1\s +i \w_1 \tau} \ ,\\
Z_2&=r_2 e^{i\varphi_2}=X_3+iX_4 = L_2\ e^{i k_2\s +i \w _2\tau} \ ,
\\
t&=\kappa \tau\ , \\ y&=mR\s + q\tau \ .
\label{claschi}
\end{aligned}
\ee
where $0\leq \s < 2\pi $ and $m,\ k_1,\ k_2$ are integer numbers.
When $k_1=k_2=1$, these solutions represent the states with maximum angular momentum having $S_{1R}+S_{2R}=N_R+1$.

In section 2.3, we will  show that these solutions  are indeed supersymmetric.
Classically, the condition~(\ref{qcond}) does not appear, since it arises due to normal ordering terms.
The classical string description applies when $N_R, J_R\gg 1$ and $S_{1,2L}=-1$ becomes negligible.

In this subsection we will reproduce the energy (\ref{eebb}) of the quantum string states for 
the cases of the circular BPS (``chiral") string~(\ref{claschi}).
The Polyakov action is given by
\be
\begin{aligned}
S &= - {1\over 4\pi \a'} \int \!\! d\s d\tau \Big( -\p_\a t\p_\a t +\p_\a y\p_\a y +r_1^2
\big(\p_\a \varphi_1 +B_1 \p_\a y\big)\big(\p_\a \varphi_1 +B_1 \p_\a y\big)
\\
&\qquad + r_2^2
\big(\p_\a \varphi_2 +B_2 \p_\a y\big)\big(\p_\a \varphi_2 +B_2 \p_\a y\big)\Big)\ .
\end{aligned}
\ee
Here the indices $\alpha $ are contracted with the world-sheet metric $h_{\alpha\beta}={\rm diag}(-1,1)$.
Then the string equations of motion are automatically satisfied provided
\be
 (k_a + B_a m R)=\pm  (\w_a + B_a q )\ ,\qquad a=1,2 \ ,
\label{qqq}
\ee
which follow from the $r_a$ equations of motion assuming that both $L_1, L_2$ are different from zero.

In addition, from the Virasoro constraints, we get the following conditions
\begin{gather}
mR q + \sum_{a=1}^2 L_a^2 (k_a + B_a m R)(\w_a + B_a q )=0 \ , \\
\kappa^2 = m^2R^2 +q^2 + \sum_{a=1}^2 L_a^2 \Big( (k_a + B_a m R)^2 + (\w_a + B_a q )^2 \Big) \ .
\end{gather}
Combining these equations, we find
\be
\kappa^2 =  (mR\pm q )^2 + \sum_{a=1}^2 L_a^2 \Big( (k_a + B_a m R) \pm (\w_a + B_a q )\Big)^{\!\!2}\ .
\label{kap}
\ee
Using~(\ref{qqq}), this gives
\be
\kappa^2 = (mR\pm q )^2\ ,
\label{trew}
\ee
where the plus sign holds for the solution with $(k_a + B_a m R)=-(\w_a + B_a q )$ (the ``Right" circular string)
whereas the minus sign holds for the solution with $(k_a + B_a m R)=(\w_a + B_a q )$ (the ``Left" circular string).

Now we would like to express the energy in terms of the physical conserved quantum numbers
$J_1, \ J_2$ and $m, n$.
We have 
\be
\begin{aligned}
E &= \int_0^{2\pi}d\s \frac{\delta S}{\delta (\partial_\tau t)} =  - {1\over\a' } \ \kappa \ ,
\\
J_a &= \int_0^{2\pi}d\s \frac{\delta S}{\delta (\partial_\tau \varphi_a)} = {L_a^2\over\a' } ( \w_a + B_a q) \ , 
\\
{n\over R} &= \int_0^{2\pi}d\s \frac{\delta S}{\delta (\partial_\tau y)} = {1\over\a' }  \Big( q + L_1^2 B_1 (\w_1 + B_1 q )+  L_2^2 B_2 (\w_2 + B_2 q )\Big) \ .
\end{aligned}
\label{varie}
\ee
Using (\ref{trew}) we find
\be
E= { |m|R\over \a' } + \left|{n\over R}- B_1J_1-B_2 J_2\right| \ ,
\label{mmc}
\ee 
reproducing exactly the result~(\ref{eebb}) of the quantum string spectrum, upon setting $B_1=B_2$.

As mentioned in the previous subsection, for some parameters, there are states that have macroscopic features but have microscopic energy. Take for instance a string in the plane 1 that is $L_2=J_2=0$ with
$m=1$, $n$ large and $mRB_1=1-1/n$, $m,n>0$ (inside the periodicity interval). 
Choosing $k_{1}=-1$ and 
using (\ref{qqq}) and (\ref{varie}) we get $J_1=mn$ and 
\be
E= {m R\over \a' }+{n\over R}\big| 1-B_1mR \big| = {mR\over\a'}+{1\over R}\ .
\label{ddp}
\ee
The energy is microscopic for $R\sim\sqrt{\a'}$ but the string is large 
in the non compact space: from (\ref{varie}) one finds
\be
L_1^2= {\a'\big|J_1 \big|\ov \big|k_1+mRB_1 \big|}={ J_1^2\a' \over \big| (mn -mRB_1 J_1)\big|} 
\label{L2}
\ee
which gives $L_{1}^2 = m n^2\a' \gg\a' $ for this state. In a coordinate system
$\varphi_1' =\varphi_1 +y/mR$ the string state that has energy (\ref{ddp}) corresponds to 
a small string with orbital angular momentum in a large orbit.

More generally, for any BPS state one can choose coordinates $\varphi_1' =\varphi_1 - k_1 y/mR$ so that 
$B_1'=B_1+k_1/(Rm)$ and the new classical solution is represented by the ansatz 
$\varphi_1'=\omega_1' \tau$, therefore $k_1'=0$, 
with the same $J_1'=J_1$, $m'=m$ and zero momentum along $y$, i.e. $n'=0$.
It represents a string extended only in the compact
dimension $y$, and looking like a particle on a large cyclotron orbit with
radius $L_1$ in the non-compact space.
In this  solution, instead of a large size we get a large
radius of the orbit. This is in agreement with the periodicity of the spectrum mentioned in 
Sect. 2.1.1.

Later we will be interested in the case of Dp branes in a background with a RR $F_{p+2}$ flux 
where the full quantum spectrum is not known
and it is not clear whether the spectrum can have any periodicity.

%For a state with $J_1=mn$ the energy (\ref{mmc}) matches (\ref{ddp}) if $n=2$.
%The size is then $L_1^2=\a' mn^2/(n-1)= 4 m\a' $.

%Note that the string moves: $\varphi_1+B_1y=1/n(\tau+\sigma)$.
%Therefore the non-compact components of the target-space energy momentum tensor 
%are non zero and gravity can see that the string is large.
%In the limit $B_1\to 1/(mR)$ the frequency of the motion goes to zero
%and the state is expected to becoming equivalent to the state $m=1,n=0,J_{1,2}=0$.

%%%%%%%%%%%%%%%%%%%%%%%%%%%%
\subsection{Supersymmetry}
%%%%%%%%%%%%%%%%%%%%%%%%%%%%%

\def\dd{\mathrm{d}}

 In this section we will prove that the circular strings (\ref{claschi}) 
with $B_1=B_2$ preserve a fraction 1/4 of the 16 supersymmetries of the background (\ref{polar}).

As shown in \cite{RTSS}, the background with $B_1=B_2$ preserves 16 supersymmetries
satisfying the condition
\be
(1- \Gamma_{1234})\varepsilon =0 \ ,
\ee
where $\Gamma_{1234}=\Gamma_1 \Gamma_2\Gamma_3 \Gamma_4$ and $\Gamma_\mu $
are the ten-dimensional Dirac matrices, $\{ \Gamma_\mu ,\, \Gamma_\nu \} = 2 g_{\mu\nu}$. As we will see below, 
$\Gamma_{1234}=\gamma_{1234}$ where  $\gamma_\mu$ are the Minkowski space Dirac matrices, $\{\gamma_\mu,\,\gamma_\nu\} = 2\eta_{\mu\nu}$.

In type IIA theory, for the circular string (\ref{claschi}) with winding and momentum we will find in addition two conditions:
\be
\gamma_{05}\varepsilon =\mp \varepsilon\ ,\qquad \gamma_{05}\gamma_{11}\varepsilon =- \varepsilon \ .
\label{qap}
\ee
The minus or plus  sign arise for the Right or Left circular string (they are related to 
the signs in $\kappa^2 =(mR\pm q)$). 
{}From these conditions we conclude that the circular string of section 2.2 preserves 1/4 supersymmetries of the background,
leaving  four unbroken supersymmetries. The reduction of 1/4 is due to the presence of two charges, winding
(fundamental string charge) and momentum (the ``wave").

Remarkably, the conditions~(\ref{qap}) are the same conditions that one obtains in the $B_1=B_2=0$ case,
so the interaction with the magnetic background does not break any additional supersymmetry.
There is only a reduction by a factor 1/2 because the background $B_1=B_2$ itself preserves  1/2 of the 32
supersymmetries of the type IIA theory.

\medskip

\noindent 1. A classical string solution is supersymmetric if  there exist covariantly constant (Killing) spinors $\varepsilon$ such that~\cite{ortin,town}
\begin{equation}
\Gamma \varepsilon = \varepsilon \ ,
\end{equation}
where $\Gamma$ is the $\kappa$-symmetry matrix \cite{BT}, which in the type IIA theory is given by
\begin{equation}\label{eqn:susymastercondition}
\Gamma = \frac{1}{\sqrt{-\det  G_{\alpha\beta}}} \dot X^\mu {X'}^\nu \Gamma_{\mu\nu} \gamma_{11} \ ,\qquad
\Gamma_{\mu\nu} = \frac{1}{2} [\Gamma_\mu ,\, \Gamma_\nu] \ ,
\end{equation}
and $G_{\alpha\beta}=\p_\alpha X^\mu\partial_\beta X^\nu g_{\mu\nu}$.
We begin by using Cartesian coordinates. 
The matrices $\Gamma_\mu $ must satisfy the Dirac algebra 
$\{ \Gamma_\mu ,\, \Gamma_\nu \} = 2 g_{\mu\nu}$  for the metric (\ref{polar}), which in Cartesian coordinates
 becomes,
\be
\begin{aligned}
\dd s^2  &= - \dd t^2 + \dd x_s^2 + \big( 1 + B_1^2 (x_1^2+x_2^2) + B_2^2 (x_3^2+x_4^2) \big) \dd x_5^2 + \dd x_1^2 + \dd x_2^2 
+ \dd x_3^2 + \dd x_4^2 
\\
& \quad
+ 2B_1 x_1 \dd x_2 \dd x_5 - 2B_1 x_2 \dd x_1 \dd x_5  +  2B_2 x_3 \dd x_4 \dd x_5 - 2B_2 x_4 \dd x_3 \dd x_5 \ ,
\end{aligned}
\ee
where $i = 6, \ldots, 9$. 
The Dirac matrices are given by
\begin{equation}\label{eqn:susycartesiangamma}
\begin{aligned}
\Gamma_\mu& = \gamma_\mu \ , \quad\mu \neq 5 \ , \\
\Gamma_5 & = \gamma_5 + B_1 \big( x_1 \gamma_2 - x_2 \gamma_1 \big) + B_2 \big( x_3 \gamma_4 - x_4 \gamma_3 \big) \ .
\end{aligned}
\end{equation} 

Consider the solution~(\ref{claschi})  ($t\equiv X_0, \ y\equiv X_5$)
\be
\begin{aligned}
X_0 &= \kappa \tau \ , & X_5  & = m R \sigma + q \tau \ ,
\\
X_1 &= L_1 \cos(k_1\sigma + \omega_1\tau) \ , & X_2  & = L_1 \sin(k_1\sigma + \omega_1\tau) \ , 
\\
X_3 &= L_2 \cos(k_2\sigma + \omega_2\tau) \ , & X_4  & = L_2 \sin(k_2\sigma + \omega_2\tau) \ . 
\end{aligned}
\ee

We now  consider eq.~(\ref{eqn:susymastercondition}). 
Making use of the results of the previous subsection 2.2, the determinant of the induced metric can be written as
\begin{equation}
\sqrt{-\det  G_{\alpha\beta}} = m^2 R^2 +\sum_{a=1}^2 L_a^2 \big( k_a +  m R B_a\big)^{\!2} \ .
\label{induc}
\end{equation}
and
\be
\begin{aligned}
\dot X^\mu {X'}^\nu \Gamma_{\mu\nu} &= - \kappa k_1 x_2 \Gamma_{01} + \kappa k_1 x_1 \Gamma_{02} - \kappa k_2 x_4 \Gamma_{03} + \kappa k_2 x_3 \Gamma_{04} + \kappa m R \Gamma_{05} 
\\
&\quad+ \big( x_1 \Gamma_{25} - x_2 \Gamma_{15} \big) \big( \omega_1 m R - k_1 q \big)
  + \big( x_3 \Gamma_{45} - x_4 \Gamma_{35} \big) \big( \omega_2 m R - k_2 q \big) \ .
\\
&= \kappa m R  \gamma_{05} 
  + \sum_{a=1}^2 x_{2a-1} \Big\{ \kappa (k_a+m R B_a) \gamma_{0\,2a}   + (\omega_a m R - k_a q) \gamma_{2a  \,5} \Big\} 
\\ 
&\quad- \sum_{a=1}^2 x_{2a}   \Big\{ \kappa (k_a+m R B_a) \gamma_{0\,2a-1} + (\omega_a m R - k_a q) \gamma_{2a-1\,5} \Big\} \ .
\end{aligned}
\ee
Hence, the supersymmetry condition~(\ref{eqn:susymastercondition}) becomes
\begin{equation}
\Big( A \gamma_{05} - \sum_{a=1}^2 \epsilon^{i j} x_{2a-i} \big\{ P_a \gamma_{0\, 2a-j} + Q_a \gamma_{2a-j\, 5} \big\} \Big) \gamma_{11} \varepsilon = \varepsilon \ , \qquad i,j = 0,1 \ ,
 \label{sadf}
\end{equation}
where $\epsilon^{i j}$ is the complete antisymmetric Levi-Civita symbol, with $\epsilon^{01} = 1$ and
%\begin{subequations}
\be
A  = \frac{\kappa m R}{\sqrt{-\det G_{\alpha\beta}}} \ , \qquad
P_a  = \frac{\kappa (k_a + m R B_a)}{\sqrt{-\det G_{\alpha\beta}}} \ , \qquad
Q_a  = \frac{\omega_a m R - k_a q}{\sqrt{-\det G_{\alpha\beta}}} \ .
\label{rre}
\ee
%\end{subequations}
If we chose a constant spinor, $\varepsilon = \varepsilon_0$, eq.~(\ref{sadf}) implies 
%
%\begin{subequations}
\begin{align}
A \gamma_{05} \gamma_{11} \varepsilon_0 & = \varepsilon_0 \ , \label{eqn:susycond03}\\
( P_a \gamma_{0b} + Q_a \gamma_{b5} ) \gamma_{11} \varepsilon_0 & = 0 \label{eqn:susycond0b}\ , \qquad a = 1, 2 \ , \quad b = 1, \ldots, 4 \ .
\end{align}
%\end{subequations}

The first equation will leave  half of the supersymmetries  of the background if, and only if, $A = \pm 1$. Multiplying the second condition by $\gamma_{b5}$ from the left we end up with
\begin{equation}
( P_a \gamma_{05} - Q_a ) \gamma_{11} \varepsilon_0 = 0 \ ,
\end{equation}
which requires $P_a = \pm Q_a$. The classical string solution must therefore satisfy the conditions
\be
\begin{aligned}
\kappa m R &= m^2 R^2 + \sum_{a=1}^2 L_a^2 \big( k_a + B_a m R \big)^2 \ , \\
\kappa (k_a+m R B_a) &= \pm ( \omega_a m R - k_a q) \ , \qquad a = 1,2 \ .
\end{aligned}
\label{grap}
\ee
 {}From the constraints~(\ref{qqq}) and~(\ref{kap}) one can easily see that these conditions are
indeed satisfied. 
 Thus we have two conditions on the Killing spinors:
\be
\gamma_{05}\varepsilon_0 =\mp \varepsilon_0\ ,\qquad    \gamma_{05}\gamma_{11}\varepsilon_0 =-\varepsilon_0\ ,
\label{apo}
\ee
as anticipated above, showing that the classical string solution 
preserves 1/4 of the 16 supersymmetries of the $B_1=B_2$ background.
Note that conditions (\ref{grap}) are also satisfied if the string rotates only on one plane,
i.e. $L_2=0$, $\w_2=k_2= 0$.

%%%%%%%%%%%%%%%%%%%%%%%%%%%%%%%%%
%\subsection{Supersymmetry in polar coordinates}
%%%%%%%%%%%%%%%%%%%%%%

\medskip

\noindent 2. It is instructive to repeat the previous derivation in polar coordinates.
{}For simplicity, here we will consider a  string rotating only in the plane 12,
\be
t=\kappa\tau \ ,\ ~~~\r _1=r_0 \ ,\ ~~~\varphi_1=\omega\tau+k\sigma \ ,\ ~~~y=q\tau+mR\sigma \ ,
\ee
and the remaining coordinates equal to zero. 

The relevant $\Gamma$ matrices are found to be 
\be
\Gamma_0=\gamma_0~, \ ~~~\Gamma_{\varphi}=r_0  \gamma_1 ~, \ ~~ ~\Gamma_y=\gamma_5+\r_0  B\gamma_1\ ,
\ee
in fact $\{\Gamma_\varphi,\Gamma_\varphi\}=2r_0 ^2$,   $\{\Gamma_\varphi,\Gamma_y\}=2r_0 ^2B$, 
$\{\Gamma_y,\Gamma_y\}=2(1+r_0 ^2B^2)$.
\vskip0.1cm

We get the condition
\be
\big(r_0 P_{1} \gamma_0\gamma_1+A \gamma_0\gamma_5+ r_0 Q_{1}\gamma_1\gamma_5\big)
\gamma_{11}\varepsilon=\varepsilon \ ,
\ee
where $P_1, Q_1, A$ were given in (\ref{rre}), (\ref{induc}),  now with $(L_1,L_2)=(r_0,0)$.  
%This  implies
%\bea
%A_{05}\gamma_{05}\gamma_{11}\varepsilon=\varepsilon  \ ,\qquad
%{A_{01}\ov A_{15}}\gamma_{05}\varepsilon=-\varepsilon \ ,
%\eea
This leads to the same conditions (\ref{apo}) as in the previous derivation.

%%%%%%%%%%%%%%%%%%%%%%%%%%%%%%%%%%%%%%%%%%%%%%%%%%%%%%%%%%%%%%%%%%%%%%%%%
\section{Magnetic field from Kaluza-Klein reduction from $B_{\mu\nu}$}
%%%%%%%%%%%%%%%%%%%%%%%%%%%%%%%%%%%%%%%%%%%%%%%%%%%%%%%%%%%%%%%%%%%%%%%%%

\setcounter{equation}{0}

%%%%%%%%%%%%
\subsection{The quantum string spectrum}
%%%%%%%%%%%%

The model is obtained by a T-dual transformation in the $y$ direction from the previous model~(\ref{polar}).
We recall the standard rules of T-duality \cite{Buscher}:
\be
g_{yy}'=g_{yy}^{-1}\ ,\qquad e^{2\phi'}= {e^{2\phi}\over g_{yy}}\ ,\qquad B'_{y\mu}={g_{y\mu}\over g_{yy}}\  ,
\qquad g'_{y\mu}={B_{y\mu}\over g_{yy}}\ .
\ee
We get (dropping  ``primes")
\be
\begin{gathered}
ds^2 = -dt^2+dx_s^2+ dr^2_1+dr_2^2 +r_1^2d\varphi_1^2+ r_2^2d\varphi_2^2
+\Lambda^{-1} 
\Big( dy^2- \big(B_1 r_1^2 d\varphi_1 + B_2 r_2^2 d\varphi_2\big)^{\!2}\Big)\ , \\
e^{2(\phi-\phi_0)}= \Lambda^{-1} \ ,\qquad 
{\rm B}_2 = \Lambda^{-1} (B_1 r_1^2 d\varphi_1+B_2 r_2^2d\varphi_2 ) \wedge dy \ , \\
\Lambda = 1+ B_1^2 r_1^2 + B_2^2 r_2^2 \ .
\end{gathered}\label{bbmet}
\ee
It represents an exact solution of string theory to all $\a'$ orders 
(being related by T-duality to a flat spacetime).

The string spectrum is obtained from the spectrum of the previous model by
exchanging $m$ and $n$ and $R\to \a'/R$. Thus
\be
\begin{aligned}
\a ' M^2 &= 2(N_L+N_R) +\a' \left({m  R\over \a'}- B_1 J_1 -B_2 J_2\right)^{\!\!2}+ {n^2\over R^2 } \\
&\quad - 2\a' B_1{n\over R} (J_{1R}-J_{1L})- 2\a' B_2{n\over R} (J_{2R}-J_{2L}) \ , \\
N_R- N_L & = m n\ .
\label{specB}
\end{aligned}
\ee
Now the spectrum is periodic in the parameters $\gamma_{1,2} \equiv \a' B_{1,2}{n\over R}$ with period 1.

The state dual to~(\ref{tqq}) has zero winding $m=0$ and $n=1$ and 
 becomes tachyonic for  $B_1-B_2>1/(2R)$. Note that this is a Kaluza-Klein state
 of the supergravity multiplet. This tachyon was studied in \cite{RTinstab,David,RTSS}.

Consider the supersymmetric model $B_1=B_2\equiv B$. The BPS states have similar quantum numbers as in the previous T-dual case:
\be
N_L=0\ ,\qquad N_R=mn\ ,\qquad (S_{1L},S_{2L})=(-1,0)\ \ {\rm or} \ \ (0,-1) \ ,
\label{qbcond}
\ee
and mass given by (cf. eq.~(\ref{eebb}))
\be
M_{\rm BPS}= \left| {n\over R} \right| + \left|  {mR\over\a'} - B(J_1+J_2)\right| \, .
\label{eebbmu}
\ee
$B$ is now restricted to be in the interval
$0<\a' B {n\over R}<1$, outside which the spectrum is repeated periodically. For the states with maximum spin $S_{1R}+S_{2R}=N_R+1$ and $l_{1,2R}=0$ we get
\be
M_{\rm BPS}= \left| {n\over R}\right| +\left|  {mR\over\a'} - Bnm \right| \, ,
\ee
The minimum mass is achieved for $\a' {n\over R}B  \to 1$, where
the mass becomes 
\be
M_{\rm BPS}\to \left| {n\over R}\right| \, .
\ee
The macroscopic strings becoming light are now strings with large $m$ and small $n$. We will see that their size is much greater than the string length $\sqrt{\a ' }$.

%%%%%%%%%%%%%%%%%%%%%%%%%%%%%%%%%%%%%%%%%%%%%%%%%%%%%%%%%%%%%%%%%%%%%%%%%
\subsection{Classical solution for the  BPS  string using Polyakov formalism }
%%%%%%%%%%%%%%%%%%%%%%%%%%%%%%%%%%%%%%%%%%%%%%%%%%%%%%%%%%%%%%%%%%%%%%%%%

We consider a string rotating on one plane 12 only, and set $r_2=\varphi_2=0$.
The Polyakov action becomes
\be
S={1\ov 2\pi } \int\!\! d\tau d\sigma \ L \ ,
\ee
with
\be
L={1\ov 2\a' } \left( -{\dot t}^2 + {\dot r_1}^2 - { r_1'}^2+
{1\ov 1+B_1^2\r_1 ^2}\Big(\r_1 ^2({\dot\varphi_1}^2-{\varphi_1'}^2)+
{\dot y}^2-{ y' }^2\Big)+{2 B_1\r_1 ^2\ov 1+B_1^2\r_1 ^2}\Big({\dot\varphi_1}{y'}-
{\varphi_1'}{\dot y}\Big) \right) \, ,
\ee
where $\dot x\equiv{\p x\ov\p\tau}$ and $\ x'\equiv{\p x\ov\p\sigma}$.

The ansatz for the circular string is:
\be
\begin{aligned}
\dot t& =\kappa\ , & 
\dot \r_1 &=\r_1 '=0 ~\to~\r_1 =r_0\ , & 
\dot\varphi_1 &=\omega\ ,&
\varphi_1'&=k\ ,& 
\dot y&=q\ , &
 y'=mR\ .
\end{aligned}
\ee
The equation of motion for $\varphi_1, ~y$ are automatically satisfied and the equation
for $\r_1 $ gives
\be
{\p L\ov\p \r_1 ^2}\bigg|_{\r_1 =r_0}=0~~\to ~~(\omega+B_1mR)^2=(k+B_1 q)^2 \ .
\label{equ}
\ee
The conserved (quantized) momenta are
\be
\begin{aligned}
\a' J_1 & ={\p L\ov\p \omega}= {r_0^2\ov 1+B_1^2r_0^2}(\omega+B_1 m R) \ , \\
\a' {n\ov R} &={\p L\ov\p q}=  q - {B_1r_0^2\ov 1+B_1^2r_0^2}(k+B_1q) \ , 
\end{aligned}
\label{hhj}
\ee
giving
\be
\begin{aligned}
r_0^2\omega & =\a' {J_1}(1+B_1^2r_0^2)-B_1r_0^2 mR \ , \\
q           & =\a' {n\ov R}(1+B_1^2r_0^2)+r_0^2B_1k \ .
\end{aligned}
\ee
From the Virasoro constraint, we have $r_0^2\omega k+qmR=0$, which becomes
\be
%r_0^2\omega k+qmR=0 ~~~~\rightarrow~~~
J_1k+nm=0\ ,
\ee
we get
\be
\begin{aligned}
q^2+r_0^2\omega^2&=m^2R^2\left(1+{n^2\ov R^2}{r_0^2\ov J_1^2}\right)D^2 \ , 
&\quad\ D^2 & \equiv{1\ov r_0^2}\left[B_1\left(1-\a' {B_1J_1\ov mR}\right)r_0^2-\a' {J_1\ov mR }\right]^{2} \ , \\
m^2R^2+r_0^2k^2&=m^2R^2 \left(1+{n^2\ov R^2}{r_0^2\ov J_1^2}\right) \ .
\label{D}
\end{aligned}
\ee
The remaining Virasoro constraint gives the energy, which takes the form
\be
E^2={\kappa^2\over {\a'}^2} ={1\ov 1+B_1^2r_0^2}{m^2R^2\over {\a'}^2} \left(1+{n^2\ov R^2}{r_0^2\ov J_1^2}\right)(1+D^2) \ .
\label{energypoly}
\ee
From the equations~(\ref{equ}),~(\ref{hhj}) we get
\be
 {J_1^2\ov r_0^4}={1\over{\a'}^2}\left(k- \a' B_1{n\ov R}\right)^{\!\!2}  = {n^2\ov R^2}\left({m R\ov \a' J_1}-B_1\right)^{\!\!2}
\label{r0}\ ,
\ee
which gives $r_0$ in terms of the (quantized) quantum numbers and $B_1$ (compare with (\ref{L2})):
\be
r_0^2={\a' \big|J_1 \big|\ov\big|k_1-\a' {n\ov R}B_1\big|}= 
{ J_1^2\over \big|{n\over R} ({mR\over\a'} -B_1 J_1)\big|} 
\ee
 Substituting this value we get
\be
E= \left|{n\ov R}\right|+\left|{m R\over \a' }-B_1J_1\right| \ .
\label{enresult}
\ee
This reproduces exactly the energy formula~(\ref{eebbmu}) of the quantum string spectrum for $J_2=0$.

%%%%%%%%%%%%%%%%%%%%%%%%%
\subsection{Classical solution for the  BPS  string using Nambu-Goto formalism }
%%%%%%%%%%%%%%%%%%%%%%%%%

Since we will be later interested in studying Dp branes, whose dynamics is governed by the Dirac-Born-Infeld (DBI) action, it is useful to
reproduce the previous result in the  Nambu-Goto formalism. 
%In the case of the D1 string, the DBI action in the S-dual background 
% is the same as the Nambu-Goto action
%for the F string, as implied by the rules of S-duality. 
We will see that the equations that we find for the Dp brane analog
of the rotating circular BPS string are a simple generalization of the treatment described below.

\medskip

The Nambu-Goto Lagrangian is given by
\be
L={1\over\a' } \sqrt {- \det G_{\alpha\beta} }+
{1\over\a' } B_{\mu\nu} \epsilon^{\alpha\beta}\p_\a X^\mu \p_\beta X^\nu\ ,\qquad 
G_{\alpha\beta}=g_{\mu\nu} \p_\alpha X^\mu \p_\beta X^\nu\ , 
\ee
where $\alpha,\beta =\sigma ,\tau$.

We will work in the gauge $G_{\sigma\tau }=0$. By taking the same anstatz as in section 3.2 with $r_1=r_0$ and $r_2=0$ 
(i.e. the string rotating in one plane one only)
we get
\be
L=
 {1\ov\a' \Lambda}\sqrt{(\kappa^2\Lambda - q^2-r_0^2\omega^2)
(r_0^2 k^2 + m^2R^2)}+ { B_1\ov \a' \Lambda} r_0^2 (\omega m R- q k)\ ,
\ee
with $\Lambda= 1+ B_1^2 r_0^2$.
The energy, angular momentum $J_1$ in the plane $r_1,\varphi_1 $ and the linear momentum in $y$ are obtained by
\be
\begin{aligned}
E &= %{\delta L\over \delta \dot X^0} =
{\partial L\over \partial \kappa}= {1\over\a'} \sqrt{ r_0^2 k^2 + m^2 R^2\over \Lambda - U} \ , \\
J_1 &= %{\delta L\over \delta \dot \varphi_1} =
{\partial L\over \partial \w} = - { r_0^2\w \over \a' \kappa \Lambda } 
\sqrt{ r_0^2 k^2 + m^2 R^2\over \Lambda - U}+ { B_1 r_0^2 mR\over \a' \Lambda } \ , \\
{n\over R} &= %{\delta L\over \delta \dot y } =
{\partial L\over \partial q}= - {  q \over \a' \kappa \Lambda }  
\sqrt{ r_0^2 k^2 + m^2 R^2\over \Lambda - U} - { B_1 r_0^2 k\over \a' \Lambda } \ ,
\end{aligned}
\label{uno}
\ee
where
\be
U=  {q^2 + r_0^2\w^2 \over \kappa^2} \ .
\ee
The constraint equation  $G_{\sigma\tau }=0$ becomes
\be
k J_1+nm=0\ .
\ee
The equations~(\ref{uno}) can be combined giving
\be
{U\ov\Lambda-U}={1\ov r_0^2} \left[ B_1r_0^2 \left( 1-{\a' B_1J_1\ov mR}\right) -{\a' J_1\ov mR}\right]^{2}
%({\a' \Lambda })^2{1\ov A}\bigg( \Big({n\over R} - { B_1 r_0^2 k\over \a'\Lambda } \Big)^2+
%\Big( {J_1\over r_0}+ { B_1 r_0 mR\over \a'\Lambda } \Big)^2\bigg)
\equiv D^2\ ,
\ee
or
\be
U=\Lambda{D^2\ov 1+D^2}\ .
\ee
Note that $D^2$ is the same quantity seen 
in the previous subsection eq.(\ref{D}).
The  energy square of the state becomes the same expression  
eq.(\ref{energypoly}), which we now write exhibiting the explicit $r_0$ dependence.
\be
 E^2=  
 { m^2 R^2\over {\a'}^2 } {1 + {n^2\over R^2}  {r_0^2\over J_1^2}\over 1+B_1^2 r_0^2} \left(
1+ \left[ {B_1\over r_0^2}(1- {\a' B_1 J_1\over  m R} ) - {\a' J_1\over  m R r_0^4} \right]^2\right)\ .
\label{dos}
\ee
The Hamiltonian that arises after the gauge choice $X_0=\kappa \tau $ 
is (after substituting for $q,\w$ their expressions in terms of ${n\ov R},J_1$ )
\be
H\big( {n/ R},J_1,r_0\big) ={n\ov R}{\p y\ov\p\tau}+J_1{\p\varphi_1\ov\p\tau}-L={n\ov R}q+J_1\omega-L\ .
\ee
Due to $\tau$-scaling invariance of the Lagrangian we have $H=-E$. 
Therefore the equation for $r_0$ is ${\p E\ov\p r_0}=0$ and this is seen to give the 
same eq.(\ref{r0}) as in the previous subsection. 
Therefore
\be
E=   \left|{n\over R}\right| + \left| {mR\over\a' } - B_1J_1\right| \ .
\ee 
In this way we recover the  result (\ref{enresult}) found by using the Polyakov formalism.

\medskip

The classical description reproduces the energy of the quantum string spectrum for large quantum numbers 
also in non-supersymmetric
configurations. As an example, in the appendix we compute the classical energy of a rotating folded string.

%%%%%%%%%%%%%%%%%%%%%%%%%%%%%%%%%%%%%%%%%%%%%%%%%%%%%%%%%%%%%%%%%%%%%%%%%
\section{Dp-branes interacting with a magnetic RR flux}
%%%%%%%%%%%%%%%%%%%%%%%%%%%%%%%%%%%%%%%%%%%%%%%%%%%%%%%%%%%%%%%%%%%%%%%%%

\setcounter{equation}{0}

%We perform a S-duality passing from a F-string to a D-string interacting with a $A_{\mu\nu}$  $RR$ field.
%The metric and the dilaton are obtained by the standard rules
%\be
%ds_E^2\equiv e^{-\phi/2}ds^2\ , ~~ d{s_E'}^2=ds_E^2\ ,~~\phi'=-\phi~~
%\rightarrow d{s'}^2=e^{\phi'}ds^2
%\ee

We consider the S-dual background to~(\ref{bbmet}), which is given by 
\be
\begin{aligned}
ds^2 &=  \Lambda^{1\over 2} \big( -dt^2+dx_s^2+ dr^2_1+dr_2^2+r_1^2d\varphi_1^2+ r_2^2d\varphi_2^2\big)
\ ,\\
&\quad+ \Lambda^{-{1\over 2}} 
\big( dy^2- (B_1 r_1^2 d\varphi_1 + B_2 r_2^2 d\varphi_2)^2\big)  
\ ,\\
e^{2(\phi-\phi_0)}&= \Lambda \ , 
\qquad %\\
 A_2 %& 
= e^{-\phi_0}  \Lambda^{-1} 
(B_1 r_1^2 d\varphi_1+B_2 r_2^2d\varphi_2 ) \wedge dy \ .
\end{aligned}
\ee
where $\Lambda = 1+ B_1^2 r_1^2 + B_2^2 r_2^2$. This represents a solution to the classical string equations 
to the leading order in $\alpha' $ .
This background contains a flux which couples to a D string.
In order to obtain magnetic flux backgrounds for the Dp brane, we perform  T-duality transformations
on $x_s $ coordinates. Using the usual rules (given in \cite{Ortin}) one finds
\be
\begin{aligned}
ds^2 &= \Lambda^{1\over 2} \big( -dt^2+dx_s^2+ dr^2_1+dr_2^2+r_1^2d\varphi_1^2+ r_2^2d\varphi_2^2\big) \\
&\quad+ \Lambda^{- {1\over 2} } 
\big( dy_1^2+...+ dy_p^2- (B_1 r_1^2 d\varphi_1 + B_2 r_2^2 d\varphi_2)^2\big)  \ , \\
e^{2(\phi-\phi_0)}&= \Lambda ^{3/2-p/2}\ ,\\
 A_{p+1} &= e^{-\phi_0} \Lambda^{-1} 
(B_1 r_1^2 d\varphi_1+B_2 r_2^2d\varphi_2 ) \wedge dy_1\wedge dy_2\wedge ... \wedge dy_{p} \ .
\label{vbn}
\end{aligned}
\ee
Here $s=p+5,...,9$.
\smallskip

{}For simplicity, here we consider a Dp-brane with $p\geq 1$ which rotates only on the 12 plane, lying at $r_2=0$.
In this case the dependence on $B_2$ disappears.
The projection of this Dp brane on the plane 12
 describes a circle with radius 
$r_1=r_0$. The Dp brane also moves and winds on a $p-$dimensional torus in the compact space.
The compact coordinates $y_i$ have periodicity $y_i\sim y_i+2\pi R_i$, $i=1,\cdots, p$. It is convenient to  
formally describe the trajectory in the non-compact space $r_1,\varphi_1$ in terms of another circular coordinate $y_0=r_0 \varphi_1$
such that $y_0\sim y_0+2\pi r_0$.

The Dp brane action is given by
\be
S= {1\over (2\pi )^{p} } \int \!\! d\tau\, \prod_{l=1}^p \int_0^{2\pi}\!\! d\sigma_l\  L \ ,
\ee
where for this ($r_2=0$) Dp brane the Lagrangian $L$ becomes
\be
L=\mu_p e^{-(\phi-\phi_0)} \sqrt {- \det  G_{\alpha\beta} }+{\mu B_1\ov \Lambda} r_1 ~\det   M\ ,
\ee
where
\be
\mu_p = {1\over g_s l_s^{p+1}}\ ,\qquad  g_s= e^{\phi_0}\ ,\qquad l_s= \sqrt{\a' }\ .
\ee
Here $\alpha,\beta = 0,1,\cdots ,p$ and
\be
G_{\alpha\beta}=g_{\mu\nu}{dX^\mu\ov ds_\alpha}{dX^\nu\ov ds_\beta}\ ,
\ee
with $ds_{\alpha}=(d\tau,d\sigma_1,\cdots ,d\sigma_p)$ and 
\be
M_{\alpha\beta} \equiv {dy_\beta\ov ds_\alpha} \ ,
\ee
with $y_{\beta}=(y_0,y_1,\cdots ,y_p)$. Note the dimension $[B_1]=[1/R]$.

\vskip0.5cm

We take the ansatz:  
\begin{align}
X^0& =\kappa\tau\ ,&  
r_1&=r_0\ ,& 
r_2&=0\ ,&
\varphi_1&=\omega\tau+ k_i\sigma_i\ ,& 
y_i&=q_i\tau+\tilde m_{ij}\sigma_j\ .
\end{align}
Using our compact notation, we write $y_0=q_0\tau+\tilde m_{0j}\sigma_j$ with 
$q_0=r_0 \omega$ and $\tilde m_{0j}\equiv r_0 k_j$.
Note that  $q_\alpha\equiv \dot y_\alpha $ and the momenta are $p_{y_\alpha}={\p L\ov\p \dot y_\alpha}$.
\vskip0.2cm
The constants of motions are: 

\begin{itemize}
\item $k_i$;
%: $\tilde m_{0j}=r_0k_j$ (for $y_0\sim y_0+2\pi r_0$);

\item the compact windings $m_{ij}$: $\tilde m_{ij} \equiv R_i m_{ij}$ (for $y_i\sim y_i+2\pi R_i$);

\item the momenta in the compactified directions $p_{y_i}={n_i\ov R_i}={\p L\ov\p q_i}$;

\item the angular momentum $J_1= {\p L\ov\p \omega}$ corresponding to  the ``momentum" 
$p_{y_0}={J_1\ov r_0}={\p L\ov\p q_0}$.

\end{itemize}

\noindent Note that $k_i,~m_{ij},~J_1,~n_i$ are integer numbers.
\vskip0.2cm

We take the gauge 
\be
G_{0j}=g_{\mu\nu}{dX^\mu\ov d\tau}{dX^\mu\ov d\sigma_j}=0 ~, \qquad j=1,\cdots, p \ .
\ee
This can be rewritten as $\sum\limits_{\alpha=0}^p q_\alpha \tilde m_{\alpha j}=0$ for each $j=1,\cdots, p$.
Since ${\p \det (M)\ov \p q_\alpha}\tilde m_{\alpha j}=0$ for $j=1,\cdots,p$ and ${\p \sqrt {- \det  G}\ov \p q_\alpha}\sim q_\alpha$,
this implies
\be
\sum_{\alpha=0}^p {\p L\ov\p q_\alpha} \tilde m_{\alpha j}=
\sum_{\alpha=0}^p p_{y_\alpha} \tilde m_{\alpha j}=J_1 k_j + \sum_{i=1}^p n_i m_{ij}=0\ ,\qquad j=1,\cdots,p \ .
\label{constr}
\ee
Another consequence is that the product of the matrix $M$ times its transpose, i.e.  $M\cdot M^T$, is block-diagonal 
and one gets $\det (M)= q \sqrt{\det (G_{ij})}$ with $q\equiv \sqrt{\sum_{\alpha=0}^p q_\alpha ^2}$. 
Also, note that $G_{00}=\kappa^2\Lambda -q^2$.
\vskip0.2cm
Therefore we can rewrite the Lagrangian in this gauge as:
\be
L={\mu_p\ov\Lambda}\sqrt{(\kappa^2\Lambda -q^2)\det (G_{ij})}  +{\mu_p B_1r_0\ov\Lambda} q \sqrt{\det (G_{ij})} \ .
\ee

Consider now, for fixed $G_{ij}$, the vector $\vec q$ whose components are the dynamical variables $q_\alpha=\dot y_\alpha$. We note that $L$ is invariant under rotations of $\vec q $ . Therefore we can take a frame where $\vec q=(q,0,\cdots,0)$ getting
\be
p_y\equiv\sqrt{\sum_{\alpha=0}^p p_{y_\alpha} ^2}={\p L\ov\p q}={-q\ov\sqrt{\kappa^2\Lambda -q^2}}{\mu_p\ov\Lambda}\sqrt{\det (G_{ij})}+
{\mu_p B_1r_0\ov\Lambda}\sqrt{\det (G_{ij})} \ ,
\ee 
from which it follows that
\be
{q^2\over \kappa^2}=\Lambda{D^2\ov 1+D^2}~\ ,\ ~~~
D^2 \equiv \left({\Lambda\ov\mu_p}{\sqrt{(J_1/r_0)^2+\sum_{i=1}^p (n_i/R_i)^2}\ov \sqrt{\det  (G_{ij})}}-B_1 r_0\right)^{\!\!2} \ .
\label{Dpp}
\ee
By using the relations~(\ref{constr}) one verifies the following identity:
\begin{align}
{\sqrt{(J_1/r_0)^2+(n/R)^2}\ov \sqrt{\det (G_{ij})}} & ={J_1/r_0\ov\tilde m}~\ , & 
\tilde m & \equiv \det  (\tilde m_{ij}) \ , &
{n\over R} & \equiv\sqrt{\sum_{i=1}^p {n_i^2\over R_i^2}} \ .
\label{id}
\end{align}
We fix $r_1 =r_0$ by requiring ${\p H\ov\p r_0}=0$ where $H$ is the Hamiltonian.
This Hamiltonian describing the dynamics of the space coordinates at fixed $\kappa$ is obtained
by substituting for $q_i,\omega $ their expressions in terms of ${n_i\ov R_i},J_1$  in the general formula
\be
H\big( {n_i\over R_i},J_1,r_0 \big) =\sum_{i=1}^p{n_i\ov R_i}{\p y_i\ov\p\tau}+J_1{\p\varphi\ov\p\tau}-L=\sum_i{n_i\ov R_i}q_i+J_1\omega-L \ .
\ee
Due to $\tau$-scaling invariance of the Lagrangian we have $H=-E$, where the energy of the state is
\be
\begin{aligned}
E&\equiv {\p L\ov\p\kappa}=\mu_p \sqrt{\det (G_{ij})\ov \Lambda -q^2/\kappa^2}=\mu_p\sqrt{\det (G_{ij})(1+D^2)\ov\Lambda} \\ 
&=\mu_p\tilde m
\sqrt{{1 + {n^2\over R^2}  {r_0^2\over J_1^2}\over 1+B_1^2 r_0^2} \bigg(
1+ \Big[ {B_1\over r_0^2}(1- {B_1 J_1\over \mu_p \td m } ) - {J_1\over \mu_p \td m  r_0^4} \Big]^2\bigg)} \ .
\end{aligned}
\label{enexpr}
\ee
In the last step we have used the definitions and the identities ~(\ref{Dpp}) and ~(\ref{id}).
\vskip0.1cm
\noindent
{}From now on, the calculation follows identical steps as in section 3.3.
We find $r_0$ by the equation ${\p E^2\ov\p r_0^2}=0$. This gives again
\be
r_0^2={J_1^2 \ov {n\ov R} |\tilde m\mu_p -B_1{J_1}|}\ .
 \label{rzer}
\ee
Substituting this value into the above expression for $E$ we get
\be
E={n\ov R}+\big|\tilde m \mu_p-B_1J_1\big| =\sqrt{\sum_{i=1}^p {n_i^2\over R_i^2}}+\big|R_1\cdot R_2\cdots R_p ~ \det ( m_{ij}) \mu_p-B_1J_1\big| \ .
\ee
In terms of the  tension of the Dp brane  $\tau_p =\mu_p/(2\pi)^p$ and the volume of the torus $T^p$
this becomes
\be
E = \sqrt{\sum_{i=1}^p {n_i^2\over R_i^2}} + \big|\tau_p\, {\rm Vol}(T^p) \, m - B_1 J_1\big| \ ,
\ee
where 
\be
m \equiv \det ( m_{ij}) ={1\over {\rm Vol}(T^p) }\int \!\! dy_1\wedge \cdots \wedge dy_p  \ .
\ee
is the winding number of the Dp brane around the $T^p$-torus. Note that
the  term $\tau_p\times {\rm Vol}(T^p) \times m$ is the expected contribution to the energy of the form ${\rm tension}\times {\rm volume}\times 
{\rm winding}$. This term is $O(1/g_s)$.

Thus we find that the energy has the same form as the energy~(\ref{enresult}) of the circular BPS string
(in the particular case $J_2=0$). This Dp brane solution is also supersymmetric as it is related
by dualities to the circular BPS string.
In the particular $p=2$ and $B_1=B_2=0$ case, it agrees with the energy formula for the analogous membrane BPS solution
found in \cite{floratosOld,BRR} (see eq. (2.42) in \cite{BRR}).

\medskip

Let us now consider some physical implications of the energy formula.
We take $R_i\sim  l_s$. In the absence of magnetic fields and assuming $g_s\ll 1$ the energy of the brane is 
essentially  given by the winding contribution
\be
E\sim {1\over g_s l_s }  m  \gg {1\over l_s} \ .
\ee
Similarly to the case of the string, 
we find that there are   Dp branes which become  light when
the magnetic flux gets to some value, $B_1 J_1 \sim \tau_p{\rm Vol}(T^p) m $. 
These are states with large $m$, and with $n_i$ of order 1.
The energy becomes
\be
E\sim \sqrt{\sum_{i=1}^p {n_i^2\over R_i^2}} \ll {1\over g_s l_s }  m  \ .
\ee
Strikingly, these Dp branes become macroscopic since $r_0$ (and in fact the proper distance) goes to infinity
when $B_1$ approaches $\tau_p{\rm Vol}(T^p) m/J_1 $ (see eq.~(\ref{rzer})).
In the absence of
magnetic fields they have size $r_0\sim {J_1\over \sqrt{m}} l_s \sqrt{g_s}$, which is typically small in the perturbative regime, but it may be large by a suitable choice of quantum numbers (satisfying of course the 
constraint (\ref{constr})).

%In addition, note that states with large $m\gg 1/g_s$ and $n_i$ of order 1 describe macroscopic 
% Dp branes in the absence of
% magnetic fields of size $r_0^2\sim J l_s^2 g_s\sim m l_s^2 g_s\gg l_s^2$. 
% The energy of these states at the critical value of the magnetic field will also become of order $l_s$.

In the case of the model (\ref{polar}), the physical string spectrum is periodic in the magnetic field parameters. A very interesting open question is whether the full quantum spectrum of Dp brane states in the background (\ref{vbn}) could also have some analogous periodicity. 

%%%%%%%%%%%%%%%
\section{Discussion}
%%%%%%%%%%%%%%%%%%%%%%

To summarize, we have computed the energy of Dp branes in the presence of magnetic RR flux backgrounds
and identified a family of BPS rotating Dp brane solutions which are invariant under four supersymmetry 
transformations. There are some potentially interesting applications.
Since the solutions are BPS, the mass formula should be  protected from  quantum string theory corrections
and therefore it should be possible to extrapolate it to strong coupling, where the gravitational field of the brane
becomes important and these branes could become black holes, analogous to the black holes of \cite{Sen}, but moving in magnetic fields.
Also, it may be possible to construct the Dp brane supergravity solution with the addition of the magnetic RR $F_{p+2}$ flux by starting with a Dp' brane background, adding magnetic parameters as in section 2 and
perform several dualities, providing a model for AdS/CFT correspondence which might exhibit some interesting effects.

We have seen that there are macroscopic string and Dp brane states which become light for some values of the magnetic field parameters $B_1, B_2$. 
In general, the presence of many light classical macroscopic states in a non-supersymmetric background could be a sign of potential instabilities. In the non-supersymmetric case, the quantum spectrum is known  \cite{RTmagnetic,RTinstab} to contain tachyons in some range of the parameters $B_1, B_2, R$ (see section 2).
Such instabilities in principle can arise both from string modes or from supergravity modes.
In the first case, like the model of section 2.1, where tachyons arise in the winding sector, the supergravity background is classically stable, but the string theory is unstable in some range of the parameters.
In the second case, like the model of section 3.1, the supergravity background itself is classically unstable
(which allows a study of tachyon instabilities in a field theory setting \cite{David}).

{}From the quantum spectrum (\ref{qqqq}) (see also appendix A), it can be seen that, when $B_1\neq B_2$, even states with 
$N_L=0$, $N_R=mn$  can become tachyonic (when $B_1=B_2$ the mass squared of these states
is manifestly positive definite).
%$N_L=0,\ N_R=mn$, 
% but rather in the sector $N_R=N_L$. 
%The classical solution describing the tachyon in the model of section 2
% is of the form $Z_1=L e^{ik\sigma}\cos (\omega\tau ),\ y=mR\sigma $. In the model of section 3 one
% has $y= q\tau $. This solution represents a pulsating circular string with $J_1=0$.
The  energy of the corresponding classical solutions never becomes imaginary because, from the Virasoro constraint,
$E^2$ is proportional to $(\dot X^i\dot X^j+  {X^i}' {X^j}')g_{ij}\geq 0$ where $g_{ij}$ is the spatial
part of the metric. The classical string may become light, but not tachyonic.

Since the presence of fluxes in string-theory backgrounds are important for moduli stabilization,
a  very interesting question is whether  instabilities could also arise for non-supersymmetric
 flux compactification models in some range of the parameters. If this is the case, this effect could constraint the number of stable vacua. A study in this direction was done in \cite{Denef2}, looking for instabilities of the supergravity background. 
%However, as pointed out above, in some cases the supergravity background can be 
% perfectly stable
% but tachyonic instabilities can appear, in particular, in excited string modes. 
%According to the studies of \cite{RTinstab,RTSS},
% in  non-supersymmetric magnetic backgrounds, the most typical situation is to have tachyonic states in
% the spectrum.

The reduction of the energy of a state originates due to the standard  gyromagnetic interaction.
This effect is universal and it is present for any quantum state with spin that moves in a magnetic field.
In string theory, the effect can be stronger due to the existence of states with arbitrarily large values of the spin,
for which the negative gyromagnetic coupling can be important even for weak magnetic fields
(for example, there are magnetic string models which become unstable for infinitesimal values of the magnetic field, see sect. 6 in  \cite{RTmagnetic}).
For strong magnetic fields, one needs to take into account $O(B^2)$ effects where gravity gets into the game.
Finding the full quantum spectrum in this case is in general highly  complicated,
but we have seen that the energy of quantum states with large spin can be obtained with a good accuracy
by studying the classical dynamics.  In non-supersymmetric backgrounds, this may signal potential 
instabilities by the presence of light states with large spin.

%Nevertheless, to
%explore the possible vacua, it may be useful to understand how the flux precisely
%affects the energy of strings and branes.

%%%%%%%%%%%%%%%%%%%%%%%%%%%%%%%%%%%%%%%%%%%%%%%%%
\section*{Acknowledgments}
%%%%%%%%%%%%%%%%%%%%%%%%%%%%%%%%%%%%%%%%%%%%%%%%%%

We would like to thank Jaume Garriga, Jaume Gomis and  Paul Townsend for useful discussions.
R.I. would like to thank the Departament ECM of U. Barcelona  where this work was carried out
for hospitality. This work is also supported by the European
EC-RTN network MRTN-CT-2004-005104. J.R. also acknowledges support by MCYT FPA
2004-04582-C02-01 and CIRIT GC 2005SGR-00564. J.L.C. is also supported by the Spanish's \textit{Ministerio de Educaci\'on y Ciencia}, FPU fellowship (ref:~AP2003-4193).

\setcounter{section}{0}
%%%%%%%%%%%%%%%%%%%%%%%
\appendix{Quantum spectrum}
%%%%%%%%%%%%%%%%%%%%%%%%%%%%%%%%%%%%
\setcounter{equation}{0}

Let us review the main points of the string quantization in the background (\ref{polar}). 
We refer to  \cite{RTmagnetic} for more details.
The coordinate $y$ satisfies the free equation $\partial_+\partial_- y=0$. 
Write $y=q\tau+mR\s+y'$, where $y'$ is single-valued and define $\gamma_{1,2}=B_{1,2}mR$, taking 
$0\leq\gamma_i<1$.
Introducing complex coordinates in the planes $1,2$:
$Z_{1,2}=X_{1,2}+iY_{1,2}=r_{1,2}e^{i\varphi'_{1,2}}$ with 
$\varphi'_{1,2}=\varphi_{1,2}+B_{1,2}y$,
one gets $Z_i=Z_{iR}(\tau+\s)+Z_{iL}(\tau-\s)$ where
\be
\sqrt{2\ov\a'}Z_{iR}(s)=ia_{i0}e^{i\gamma_i s}+i
\sum_{k=1}^\infty a_{i,k} e^{i(k+\gamma_i)s}+a_{i,-k} e^{i(-k+\gamma_i)s}\ .
\ee
$Z_{iL}(s)$ has the same expression with $a_{ik} \to\tilde a_{ik} $ and 
$k+\gamma_i\to k-\gamma_i$, and similarly 
$Z^*_i=X_i-iY_i$ in terms of $a^*_{i,k}$ (R part) and $\tilde a^*_{i,k}$ (L part).
Note that $Z_{iR,L}(\s=2\pi)=e^{i2\pi\gamma_i}Z_{iR,L}(\s=0)$.

One then introduces annihilation and creation operators by $b_{i0}\equiv\sqrt{\gamma_i\ov 2} a_{i0}~, ~
b^\dagger_{i0}\equiv\sqrt{\gamma_i\ov 2} a^*_{i0}$ and
$$
b_{ik-}\equiv\sqrt{k+\gamma_i\ov 2}a_{i,k}~,~~
b^{\dagger}_{ik-}\equiv\sqrt{k+\gamma_i\ov 2}a^*_{i,k}~,~~
b_{ik+}\equiv\sqrt{k-\gamma_i\ov 2}a^*_{i,-k}~,~~
b^{\dagger}_{ik+}\equiv\sqrt{k-\gamma_i\ov 2}a_{i,-k}\ ,
$$
such that $[b_{ik\pm},b^\dagger_{jk'\pm}]=\delta_{ij}\delta_{kk'}$ and 
similarly for the Left part. 
A similar construction holds for the fermionic coordinates. Here we will consider as an example the NS sector.
In this case, the integer $k$ is replaced by a half-integer number and therefore there is no fermionic  zero mode.

The angular momentum in the plane $i$ is $J_i=J_{iR}+J_{iL}$
with
\bea
J_{iR}&=& -{1\ov 4}\sum_k(k+\gamma_i)(a^*_{ik}a_{ik}+a_{ik}a^*_{ik})+J^{\psi}_{iR}=
-b^\dagger_{i0}b_{i0}-{1\ov 2}+\sum_{k\geq 1}(b^\dagger_{ik+}b_{ik+}-b^\dagger_{ik-}b_{ik-})
+J^{\psi}_{iR}\ ,
\non\\
J_{iL}&=& -{1\ov 4}\sum_k(k-\gamma_i)(a^*_{ik}a_{ik}+a_{ik}a^*_{ik})+J^{\psi}_{iL}=
\tilde b^\dagger_{i0}\tilde b_{i0}+{1\ov 2}+\sum_{k\geq 1}
(\tilde b^\dagger_{ik+}\tilde b_{ik+}-\tilde b^\dagger_{ik-}\tilde b_{ik-})+J^{\psi}_{iL}\ ,
\non
\eea
where $J^{\psi}_{iR,L}$ are the contributions of the fermionic coordinates in the plane $i$. 
Note that there is no fermionic analog of $b^\dagger_{i0}b_{i0}+{1\ov 2}$.
The momentum in the $y$ direction is ${n\ov R}=q+B_1J_1+B_2J_2$.

One sees that
$$
\sum_i\sum_k \Big( {k+\gamma_i\ov 2} \Big)^{\!2 }\, (a^*_{ik}a_{ik}+a_{ik}a^*_{ik})+2 N'_R
=2N_R-\sum_i 2\gamma_iJ_{iR}\ ,
$$
$$
\sum_i\sum_k \Big({k-\gamma_i\ov 2}\Big)^{\! 2}\, (\tilde a^*_{ik}\tilde a_{ik}+\tilde a_{ik}\tilde a^*_{ik})+2 N'_L
=2N_L+\sum_i 2\gamma_iJ_{iL}\ ,
$$
where $N'_{R,L}$ represent the contributions of the other coordinates and of the NS fermions
before normal ordering, and
$N_{R,L}$ are the usual flat spacetime number operators
including the contribution of every coordinate and of the NS fermions: $N_{R,L}=:N_{R,L}:-1/2$.
The GSO projection implies $N_{R,L}\geq 0$. 
From this we get
\be
\a' M^2= {(mR)^2\ov\a'}+\a'\big( {n\ov R}-B_1J_1-B_2J_2\big)^2+2(N_R+N_L)
-\sum_i2\gamma_i(J_{iR}-J_{iL})\ ,
\label{vvvv}
\ee
where $J_{iR,L}\equiv S_{iR,L}\mp (l_{iR,L} +1/2)$ and $l_{iR}=b^\dagger_{i0}b_{i0}$,
\ $l_{iL}= \tilde b^\dagger_{i0} \tilde b_{i0}$,
%$Q_i=Q_{iR}+Q_{iL}\equiv b^\dagger_{i0}b_{i0}+1/2+\tilde b^\dagger_{i0}\tilde b_{i0}+1/2$
representing the contributions of the Landau Levels in the plane $i$.
Note the bounds
\be
l_{iR,L}\geq 0~,~~~~~~|S_{1R,L}\pm S_{2R,L}|\leq N_{R,L}+1 \ .
\label{bound}
\ee
The bound for $S_{iR,L}$ can be saturated if in the plane $1$ or $2$ the fermionic modes $1/2+$ or $1/2-$ 
are excited (and no other fermionic modes) plus possibly
the modes $1+$ or $1-$ (and no  other bosonic modes). 
From (\ref{bound}) one can see that
$$
\a' M^2\geq {(mR)^2\ov\a'}+\a' \big( {n\ov R}-B_1J_1-B_2J_2\big)^2 +
2(N_R+N_L)\big( 1-\sum_i{\gamma_i\ov 2} \big)
-(\gamma_1-\gamma_2)(S_{1R}-S_{2R}-S_{1L}+S_{2L})
$$
Therefore $M^2\geq 0$ for $\gamma_1=\gamma_2$, i.e. when  $B_1=B_2$.

In general for $B_1\neq B_2$ there can be tachyons. 
In particular, consider the case $B_2=0$.
Eq. (\ref{vvvv}) can be written as
$$
\a' M^2= {(mR)^2\ov\a'}+{\a'\ov (mR)^2}(mn-\gamma_1(S_{1R}+S_{1L}))^2+2(N_R+N_L)+
2\gamma_1 ( l_{1L}+l_{1R} +1 ) -2\gamma_1(S_{1R}-S_{1L})
$$
From (\ref{bound}) one have $|S_{1R,L}|\leq N_{R,L}+1\ $.
Take, for example, $n=0$, $m=1$, $R^2=\a'$, $N_R=N_L=0$,  ~$l_{1L}=l_{1R}=0$, $S_{1R}=1$,
$S_{1L}=-1$. We get 
$$
\a'M^2=1-2\gamma_1<0\ , \qquad {\rm for}\ \  1/2<\gamma_1<1\ .
$$

Another interesting tachyonic state, appearing at $B_1\neq B_2$ in a certain range of the parameters, is a state with $N_L=0$. We take $N_R=nm,~N_L=0,~l_{1,2L}=l_{1,2R}=0,~S_{1R}=nm+1$, $S_{1L}=-1$, $S_{2R}=0$, $S_{2L}=0$, obtaining
\be
\a'M^2=\Big( {mR\ov \sqrt{\a' }}+{\sqrt{\a'}n\ov R}(1-\gamma_1)\Big)^2-2(\gamma_1-\gamma_2)\ .
\label{kjh}
\ee
It follows that $M^2$ can be negative only if
$$
(B_1-B_2)>{mR\over 2\a' }\ .
$$
Setting, for instance,  $m=1$, $\gamma_1=1-1/n_0,\ B_2=0, \ R=\sqrt{\a' }$, one finds
that $M^2$ becomes negative for  $n_0> 1+n+ \sqrt{(n+1)^2+n^2}$.

Note the difference with the true BPS case which occurs for $B_1=B_2$, where $M^2$ in (\ref{kjh}) becomes a perfect square.

%\setcounter{section}{0}
%%%%%%%%%%%%%%%%%%%%%%%
\appendix{Rotating folded string}
%%%%%%%%%%%%%%%%%%%%%%%%%%%%%%%%%%%%
\setcounter{equation}{0}

Another class of widely studied quantum string states represents strings which are folded and rotate in the non-compact plane
$\r_1 ,\varphi_1$. In addition, we will consider the case when it is wrapped in the compact dimension $y$
with winding $m\geq 0$ and zero momentum $n$. 
The corresponding ansatz is 
\be
t=\kappa\tau~\ ,~~~~\r_1 =\r_1 (\sigma)~\ ,~~~\varphi_1=\omega\tau~\ ,~~~y= m R\sigma\ ,
\label{ffd}
\ee
and $\r_2=\varphi_2=0$. We will consider this string in the model (\ref{bbmet}).
It corresponds to a quantum state with
\be
N_R=N_L=N\ ,\qquad J_{1R}=J_{1L}=N\ ,\qquad J_{2R}=J_{2L}=0\ ,
\ee
which gives 
\be
\a' M^2 =4 N +\a' \left({mR\over\a' } - B_1 J_1 \right)^2 \ ,
\label{mfd}
\ee
where we assume $B_1>0$.
We will now show that the classical energy of the solution~(\ref{ffd}) reproduces this formula.

The Nambu-Goto Action is $S={1\over 2\pi}\int d\tau d\sigma L$
where the Lagrangian $L$ is 
\be
L={1\ov \a' \Lambda}\sqrt{(\kappa^2\Lambda -\r_1 ^2\omega^2)\big((d\r_1 /d\sigma)^2\Lambda+ m^2R^2\big)}
+ {B_1\r_1 ^2\ov \a' \Lambda}\omega m R \ .
\ee
%We can put $\kappa=1$ by rescaling $\tau$.
We can formally consider $\sigma$ to play the  role of ``time" and solve for $\r_1 (\sigma)$ 
by using the Hamiltonian formalism, where we define a ``conjugate momentum" and the ``Hamiltonian" by
\be
p(\sigma)=\alpha' \, {\p L\ov\p (d\r_1 /d\sigma)}~, ~~~~H(p,\r_1 )=\frac{1}{\alpha'} \, p \, \frac{d\r_1}{d\sigma}-L \ .
\ee
We  call for short:
\begin{align}
f & \equiv{\kappa^2+(\kappa^2 B_1^2-\omega^2)\r_1 ^2\ov 1+B_1^2\r_1 ^2} \ ,
& g & \equiv{ m^2 R^2\ov 1+B_1^2\r_1 ^2} \ ,
& A & \equiv -\alpha'\,H(p,\r_1 ) \ .
\end{align}
The important point is that $A$ is constant.
We find
\be
\begin{gathered}
p^2={(d\r_1 /d\sigma)^2\ov (d\r_1 /d\sigma)^2+g}f
\ , 
%\quad \text{from where}\ %~\Longleftrightarrow~
%\left(\frac{d\r_1}{d\sigma} \right)^{\!\!2}={p^2\ov f-p^2}g \ ,  
\\
\sqrt{fg^2\ov (d\r_1 /d\sigma)^2+g}=\sqrt{(f-p^2)g}=A-{B_1\r_1 ^2 mR\ov 1+B_1^2\r_1 ^2}\omega \ ,
\end{gathered}
\label{positive}
\ee
giving
\be
\left(\frac{d\r_1}{d\sigma} \right)^{\!\!2}=g{fg-\big(A-{B_1\r_1 ^2 m R\ov 1+B_1^2\r_1 ^2}\omega\big)^2\ov \big(A-{B_1\r_1 ^2 mR\ov 1+B_1^2\r_1 ^2}\omega\big)^2} \ .
\ee
Therefore $fg-(A-{B_1\r_1 ^2 mR\ov 1+B_1^2\r_1 ^2}\omega)^2\geq 0$, which is
satisfied if 
\be
0\leq \r_1 ^2 \leq \r_M^2 \ , \qquad \r_M^2\equiv{ \kappa^2 m^2R^2-A^2\ov (mR\omega-AB_1)^2} \ .
\label{rhoM}
\ee
%Hence, we can express $A$ in terms of $\r _M^2$. We find two roots:
%\be
%A_{\pm}={ m R\ov 1+B_1^2\r ^2}\left( -B_1\r _M^2\omega\pm\sqrt{\kappa^2+(\kappa^2 B_1^2-\omega^2)\r %_M^2}\right) \ ,
%\ee
%from which it follows that
%$0\leq\omega^2\leq\kappa^2 \, {(1+B_1^2\r _M^2)/\r _M^2}$.

Note that~(\ref{positive}) holds for arbitrary $r_1$ and therefore it implies that
\be
A\geq 0 \ , ~~ A-B_1r_1^2(mR\omega-AB_1)\geq 0 \ ,%~~\to ~~ 
\ee
so that
\be
\left|{A\ov mR\omega-AB_1}-B_1\r_1 ^2\right|={A-B_1r_1^2(mR\omega-AB_1)\ov |mR\omega-AB_1|} \ .
\label{positive2}
\ee

Some useful formulae, following from rewriting what said above, are:
\begin{gather}
p =\sqrt{f-\frac{1}{g} \left(A-{B_1 \r_1^2 mR\omega\ov 1+B_1^2\r_1 ^2}\right)^{\!\!2}} = 
%{1\ov \tilde m^2(1+B_1^2\r ^2)}
%(\tilde m^2(1+B_1^2\r ^2-\omega^2\r ^2)-(A(1+B_1^2\r ^2)+B_1\r ^2\tilde m\omega)^2)=
{|mR\omega-AB_1|\ov mR}\sqrt{\r _M^2-\r_1 ^2} \ , \\
\frac{d\r_1}{d\sigma}=\pm { mR\sqrt{\r _M^2-\r_1 ^2}\ov |{A\ov mR\omega-AB_1}-B_1\r_1 ^2|}\ .
\label{rho'}
\end{gather}

The energy of the state is
\begin{equation}
\begin{aligned}
E & ={1\over 2\pi}\int d\sigma {\p L\ov\p\kappa}=
{\kappa \over 2\pi\a'}\int d\sigma \sqrt{(d\r_1 /d\sigma)^2+g\ov f}
= {\kappa\over 2\pi\a'}\int d\sigma {d\r_1 \over d\sigma}{1\ov p} \\
 & = {2n\kappa \over \pi\a'}{mR\ov |mR\omega-AB_1|}
\int_0^{\r _M} {d\r_1 \ov\sqrt{ \r _M^2-\r_1 ^2}}= {n\kappa \over \a'}{ mR\ov |mR\omega-AB_1|} \ .
\end{aligned}
\end{equation}
Here we have used~(\ref{rho'}) with the $+$ sign when $r$ grows from $0$ to $r_M$ and
we have assumed that the string is folded $n$ times; a factor $4$ appears because
for $n=1$ any point $\r_1 $ of the string is obtained $4$ times as $\sigma$ goes from $0$ to
$2\pi$.

We have still to require, using~(\ref{rho'}) and~(\ref{positive2}),
\be\label{eqn:2pi}
\begin{aligned}
2\pi & =\int_0^{2\pi}\!\! d\sigma ={4n\ov |mR\omega-AB_1|}\int_0^{\r _M}\!\!d\r_1  ~
\left|{A-B_1r_1^2(mR\omega-AB_1) \ov mR\sqrt{\r _M^2-\r_1 ^2}}\right| \\ & =
{2\pi n\ov  m R}\left[{A\ov |mR\omega-AB_1|}-\epsilon B_1{\r _M^2\ov 2} \right] \ ,
\end{aligned}
\ee
where $\epsilon\equiv \mathrm{sign} (mR\omega-AB_1 )$.
The formula~(\ref{eqn:2pi}) together with~(\ref{rhoM}) gives
\be
{\kappa^2 m^2R^2\ov (mR\omega-AB_1)^2}=\r _M^2+{A^2\ov (mR\omega-AB_1)^2}=
\r _M^2+\left({ mR\ov n}+\epsilon{B_1\r _M^2\ov 2}\right)^{\!\!2} \ ,
\ee
and therefore
\be
E=  {n\ov \a'}\sqrt{\r _M^2+\left({mR\ov n}+\epsilon{B_1\r _M^2\ov 2}\right)^{\!\!2}}\ .
\ee

Finally, one can express $\r _M^2$ in terms of the angular momentum $J_1$:
\be
\begin{aligned}
J_1 &=  {1\over 2\pi}\int \!\!  d\sigma{\p L\ov\p \omega}
= {1\over 2\pi\a' }\int \!\!d\sigma 
\left( {-\omega\,\r_1^2\ov 1+B_1^2\r_1 ^2}\sqrt{(d\r_1 /d\sigma)^2+g\ov f}
+ {B_1 mR\r_1 ^2\ov 1+B_1^2\r_1 ^2} \right) \\
&= {4n\over 2\pi\a'|mR\omega-AB_1| }\int_0^{r_M} \!\! d r_1\ \frac{-\omega r_1^2mR+B_1r_1^2\big( A-B_1r_1^2(mR\omega-AB_1) \big)}{ (1+B_1^2r_1^2)\sqrt{r_M^2-r_1^2}} \\
&=-\epsilon {n\over 2\a'}r_M^2 \ .
\end{aligned}
\ee
Thus
\be
\a' E^2=2n|J_1|+ \a' \left({mR\over\a'} -B_1J_1\right)^{\!\!2}\ ,
\ee
in agreement with~(\ref{mfd}).

\end{document}